\documentclass[conference]{IEEEtran}
\IEEEoverridecommandlockouts
\usepackage{cite}
\usepackage{amsmath,amssymb,amsfonts}
\usepackage{algorithmic}
\usepackage{graphicx}
\usepackage{textcomp}
\usepackage{xcolor}
\def\BibTeX{{\rm B\kern-.05em{\sc i\kern-.025em b}\kern-.08em
    T\kern-.1667em\lower.7ex\hbox{E}\kern-.125emX}}
\begin{document}

\title{A Low-Field Magnetic Resonance Signal Transmission and Reception Processing Platform\\
}

\author{\IEEEauthorblockN{1\textsuperscript{st} Zesong Jiang}
\IEEEauthorblockA{\textit{Institute of Advanced Technology} \\
\textit{University of Science and Technology of China}\\
Hefei, China \\
jiangzesong@mail.ustc.edu.cn}
*Corresponding author
~\\
\and
\IEEEauthorblockN{2\textsuperscript{nd} Muhang Zhang}
\IEEEauthorblockA{\textit{Institute of Advanced Technology} \\
\textit{University of Science and Technology of China}\\
Hefei, China \\
zhangmuhan@mail.ustc.edu.cn}
~\\
\and
\IEEEauthorblockN{3\textsuperscript{rd} Qing Zhang}
\IEEEauthorblockA{\textit{School of Information Science and Technology} \\
\textit{University of Science and Technology of China}\\
Hefei, China \\
qzhiang@mail.ustc.edu.cn}
~\\
\and
\IEEEauthorblockN{4\textsuperscript{th} Yuchong Xie}
\IEEEauthorblockA{\textit{School of Information Science and Technology} \\
\textit{University of Science and Technology of China}\\
Hefei, China \\
yc\_xie@mail.ustc.edu.cn}
}

\maketitle

\begin{abstract}
In magnetic resonance imaging (MRI), the spectrometer is a fundamental component of MRI systems, responsible for the transmission of radiofrequency (RF) pulses that excite hydrogen nuclei and the subsequent acquisition of MR signals for processing. However, detailed knowledge about this component remains largely inaccessible due to the proprietary nature of commercial systems. To address this gap, we present an FPGA-based platform specifically designed for MRI signal transmission and reception in low-field MRI applications. Additionally, with appropriate chip replacements, this platform can be adapted for use in mid- and high-field MRI systems. This platform leverages Direct Digital Synthesis (DDS) technology to generate RF pulses, offering the flexibility to quickly and precisely adjust soft pulse parameters to meet the specific requirements of the MRI system. Additionally, the platform processes MRI signals through digital downconversion techniques and utilizes CIC and FIR filters to obtain baseband signals. Experimental testing of this platform has yielded promising results. We hope that this work will inspire further research and development in the field of MRI spectrometer design. Furthermore, it is worth noting that with the replacement of relevant chips, this system can also be adapted for use in mid- and high-field MRI systems.
\end{abstract}


\section{Introduction}
The functions of a magnetic resonance imaging (MRI) spectrometer can be broadly categorized into two key areas: the transmission of RF pulses and the acquisition and processing of magnetic resonance signals. Currently, most MRI systems \cite{obungoloch2023site,thoeng2023new, martin2023relaxation,parsa2023single,michal2020low,wald2020low} utilize DDS technology for the generation of RF pulses. Compared to traditional analog frequency generation, DDS offers significant advantages, including high frequency resolution, precise phase accuracy, and rapid frequency switching. Typically, DDS functionality is achieved by controlling specialized DDS chips via FPGA or DSP \cite{liang2013digital,gebhardt2016fpga}, which are responsible for generating the desired waveform. However, this approach has inherent limitations in terms of waveform generation flexibility, speed, and integration, primarily due to its dependence on external control by FPGA or DSP \cite{liu2011radiofrequency}.

In response to these limitations, this paper presents a novel implementation where DDS functionality is fully integrated within an FPGA, enabling the generation of both baseband and carrier waveforms. These waveforms are subsequently modulated to produce pulse signals, which are then converted to RF pulses via digital-to-analog (DA) conversion. This approach offers the flexibility to adjust MRI parameters rapidly, facilitates swift RF pulse generation, and conserves chip area.

Furthermore, the spectrometer within the MRI system plays a critical role in signal acquisition and processing \cite{morris2002wide}. Our design employs an eight-channel system to simultaneously capture MRI data, thereby enhancing the signal-to-noise ratio (SNR) and accelerating data acquisition. The signal processing pipeline includes digital downconversion and decimation. Digital downconversion is achieved by generating a local oscillator at a frequency matching the MRI signal, enabling efficient extraction of the baseband signal. Given the high data rates of MRI signals, decimation techniques are utilized to reduce the data rate, thereby improving the efficiency of subsequent MRI image processing \cite{villena2016fast,cordeiro2015fpga}. The decimation process involves using Cascaded Integrator-Comb(CIC) filters for decimation filtering, along with CIC compensation filters, half-band filters, and FIR filters to ensure optimal passband performance and high-quality output signals.

By integrating both transmission and reception functionalities within a single FPGA, the proposed design greatly enhances system integration, reduces platform costs, and ensures phase coherence between the local oscillator used for digital downconversion and the RF carrier signal. This integration not only improves MRI image quality but also minimizes artifacts. Additionally, a storage module is included to temporarily retain the processed magnetic resonance data, ensuring lossless signal transmission. The data is then transmitted via Ethernet to a PC, where it is reconstructed into phantom images using k-space image reconstruction.

The major contributions of this work are as follows:

\begin{itemize}
    \item We have developed the first FPGA-based platform capable of simultaneously implementing RF pulse transmission and magnetic resonance signal acquisition and processing. Compared to previous platforms that relied on multiple chips to achieve transmission or reception functions, our platform utilizes a single FPGA chip to perform both transmission and reception. This significantly improves system integration and functional integration.
\end{itemize}
\begin{itemize}
    \item The platform employs a single FPGA chip to achieve RF pulse transmission and magnetic resonance signal processing, offering a high degree of integration and flexibility. It allows for easy control of the phase difference between transmission and reception, with lower transmission latency compared to previous designs.
\end{itemize}
\begin{itemize}
    \item Testing with oscilloscopes and MRI systems has demonstrated that the platform delivers excellent FID signal quality and imaging performance, establishing it as a viable replacement for traditional spectrometers.
\end{itemize}
\section{Background}
\subsection{Radio Frequency Pulse Generator}

DDS is a technology used for generating precise digital waveforms. The architecture of DDS typically comprises modules such as the waveform lookup table, phase accumulator, and phase register. Given the abundant resources of lookup tables and registers within an FPGA, coupled with its common application in high-speed signal processing circuits, FPGA is particularly well-suited as a platform for DDS implementation.

\begin{figure}[h]
  \centering
  \includegraphics[width=0.5\textwidth]{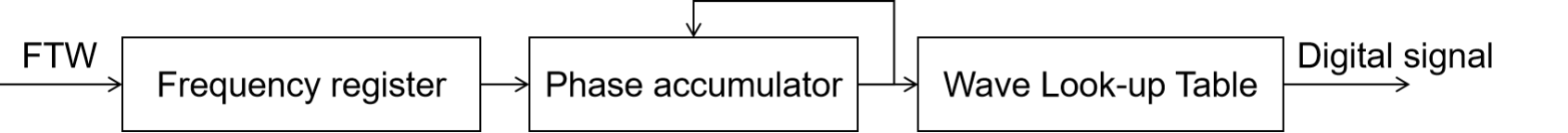}
  \caption{The Structure of DDS.}
  \label{DDS}
\end{figure}

The DDS process is illustrated in Figure \ref{DDS}. Initially, a frequency tuning word (FTW) $K$ is provided as input. During each clock cycle, the FTW $K$ is added to the phase accumulator, which generates a corresponding phase address. This address is then used to access the waveform lookup table, outputting the corresponding amplitude value. By sequentially outputting these amplitude values, a digital waveform is constructed. Subsequently, this digital waveform is converted by a DAC to produce the desired analog waveform\cite{sagcan2015fpga}.

\begin{equation}
        FTW = \text{Int}\left[\left(\frac{f_{\text{out}}}{f_{\text{clk}}}\right) \times 2^N\right]
        \label{1}
\end{equation}

The frequency of the DDS output waveform can be calculated using Equation (\ref{1}). It can be observed from Equation (\ref{1}) that the output frequency is determined solely by two factors: the sampling clock frequency $f_{\text{clk}}$ and the frequency tuning word $K$.

After generating the baseband and carrier signals using DDS, these signals must be modulated by a waveform modulator. This process aligns the carrier frequency with the resonance frequency of hydrogen nuclei, thereby converting the baseband signal into an RF signal. The modulated signal is then transformed into an analog RF pulse signal via a DAC. This RF pulse drives the RF coil, generating an RF magnetic field, which in turn excites the hydrogen nuclei within the body. The induced resonance response produces MRI signals that can be utilized for imaging purposes\cite{xiao2021rf}.

\subsection{Digital MRI Receiver}
In MRI systems, the MRI signal can be conceptualized as a narrowband signal modulated onto the Larmor precession frequency. The analog free induction decay (FID) signal \( S_{rf}(t) \) input to the ADC can be expressed as the modulation of the carrier signal \( c(t) \) onto the baseband imaging signal \( S_{bf}(t) \)\cite{dalvi2018field}:
\begin{equation}
    S_{bf}(t) = z_{bfi}(t) + jz_{bfq}(t)
    \label{eq:baseband_signal}
\end{equation}
\begin{equation}
    c(t) = \cos[w_0 t + \theta_c] + j \sin[w_0 t + \theta_c]
    \label{eq:carrier_signal}
\end{equation}
where \( w_0 \) is the resonance frequency corresponding to the selected imaging slice, \( \theta_c \) is the initial phase of the carrier signal, \( S_{bf}(t) \) is the desired baseband signal containing the complete amplitude, frequency, and phase information required for imaging, and \( z_{bfi}(t) \) and \( z_{bfq}(t) \) represent the in-phase and quadrature components of the baseband signal, respectively. The modulated imaging signal can thus be expressed as:
\begin{equation}
    S_{rf}(t) = \text{Re}\left[S_{bf}(t) \cdot c(t)\right] 
    \label{eq:modulated_signal}
\end{equation}
The objective of digital downconversion is to demodulate the baseband signal \( S_{bf}(t) \) from the modulated signal \( S_{rf}(t) \). According to the sampling theorem, after sampling and quantizing the signal at intervals \( T_s \) using the ADC, the discrete sequence \( S_{rf}[n] \) can be represented as:
\begin{equation}
    S_{rf}[n] = z_{bfi}(n) \cos[w_0 nT_s + \theta_c] - z_{bfq}(n) \sin[w_0 nT_s + \theta_c]
    \label{eq:discrete_signal}
\end{equation}

Thus, two orthogonal local oscillator sequences with the same frequency and coherent phase as the carrier \( c(t) \) are generated using the DDS method:

\begin{equation}
c_{i,q}[n] = \cos[w_0 n T_s + \theta_c]
\label{eq1}
\end{equation}

By mixing the sampled imaging signal sequence \( S_{rf}[n] \) with the local carrier, we obtain:

\begin{equation}
S_{i,q}[n] = S_{rf}[n] \cdot c_{i,q}[n]
\label{eq2}
\end{equation}

Substituting Equations (\ref{eq:discrete_signal}) and (\ref{eq1}) into the above expression yields:

\begin{equation}
S_{i,q}[n] = \frac{1}{2} \cdot z_{bfi,q}[n] - I_{rf}[n]
\end{equation}

After performing digital downconversion on the acquired magnetic resonance signals, it is necessary to design various filters for decimation and filtering processes. Due to space constraints, this discussion will focus on the CIC filter and the CIC compensator filter. CIC filters are employed to achieve significant sample rate changes in digital systems and are therefore commonly used in high sample rate scenarios. A CIC filter consists of an equal number of integrator and comb filters arranged in a cascaded configuration. By selecting an appropriate number of CIC filter stages, the frequency response of the CIC filter can be effectively tuned. Equation (\ref{eq:CIC}) provides the relevant transfer function for the CIC filter:

\begin{equation}
H(Z) = \frac{1}{1 - Z^{-1}} \cdot (1 - Z^{-D}) = H_1(Z) \cdot H_2(Z) 
\label{eq:CIC}
\end{equation}

The amplitude response of multi-stage cascaded CIC filters often exhibits passband ripple. To correct the frequency response and mitigate this ripple, a filter with an amplitude response that is the inverse of the CIC filter can be employed. These filters are referred to as CIC compensation filters\cite{xu2019design}.


\section{Methods}



\subsection{Hardware implementation}
\begin{figure}[h]
  \centering
  \includegraphics[width=0.5\textwidth]{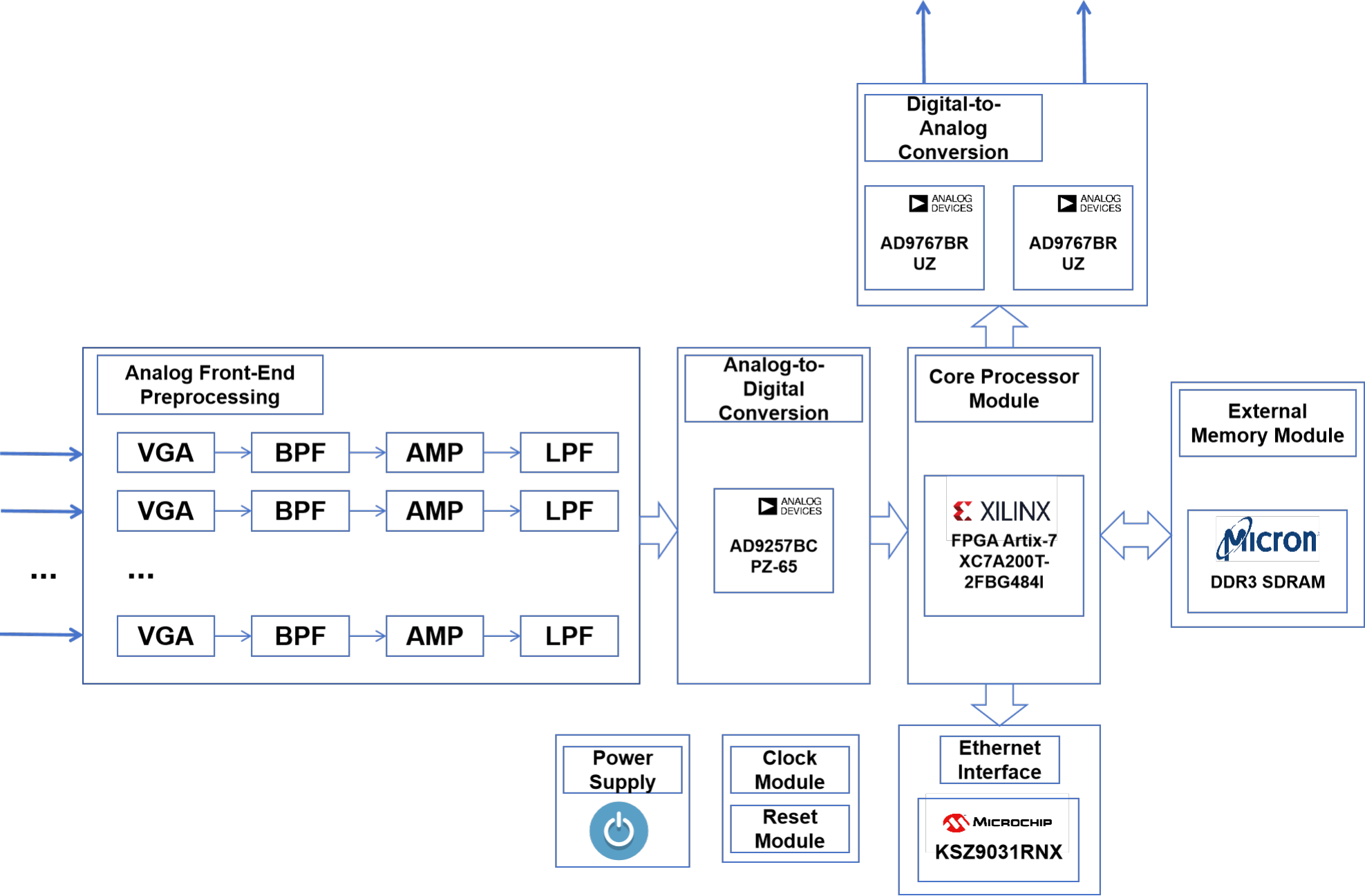}
  \caption{The Hardware Framework of the Platform}
  \label{platform}
\end{figure}

\begin{figure}[h]
  \centering
  \includegraphics[width=0.5\textwidth]{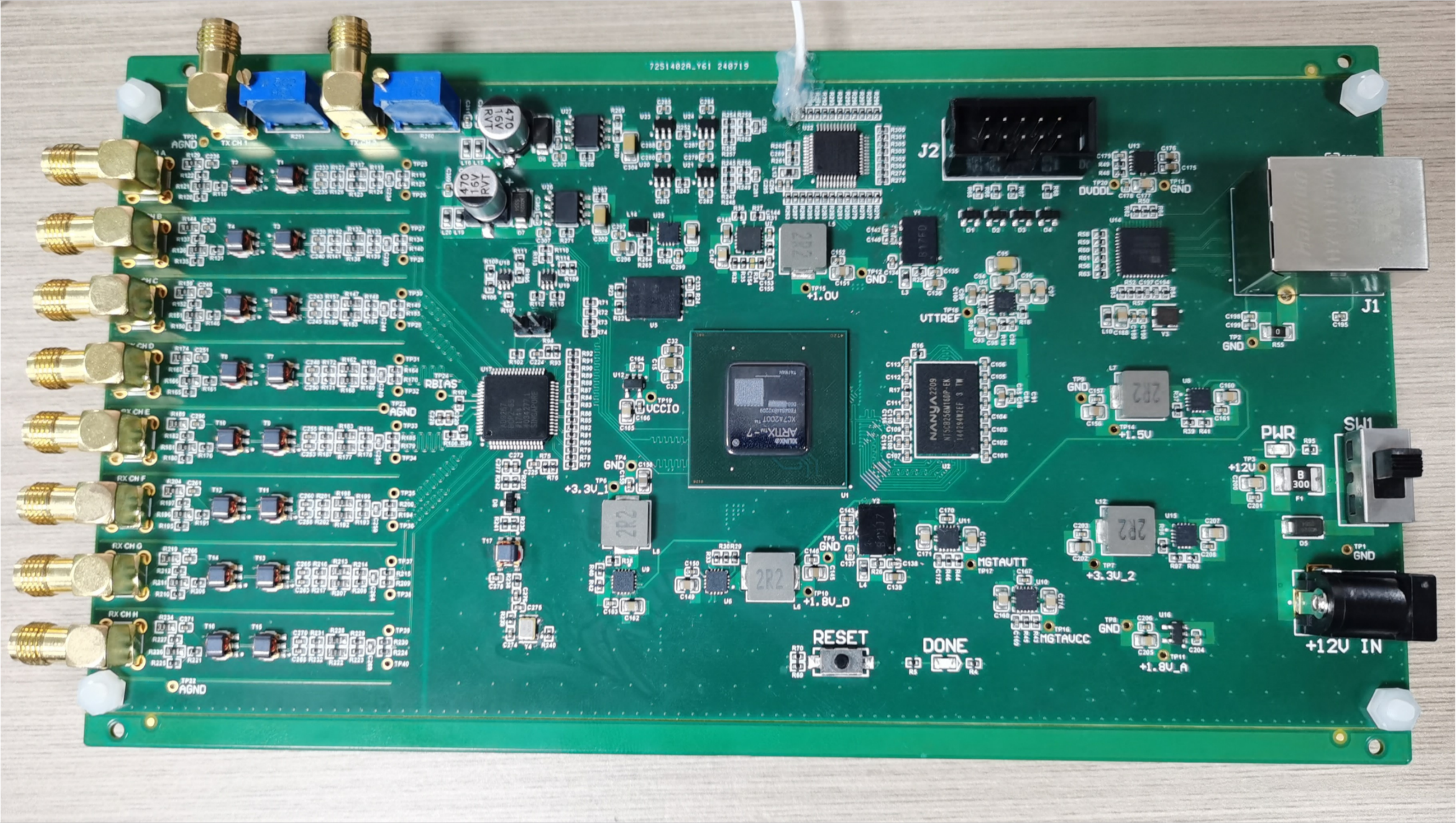}
  \caption{The Physical Diagram of the MRI Signal Transmission and Reception Processing Platform Board}
  \label{实物图}
\end{figure}

This paper introduces a comprehensive platform designed for magnetic resonance signal transmission and reception processing, specifically tailored for low-field MRI. The hardware architecture, as illustrated in Figure~\ref{platform}, comprises two primary subsystems: RF pulse generation, and magnetic resonance signal acquisition and processing. Additionally, the platform integrates a DDR3 module for intermediate data storage, an Ethernet interface for transmitting MRI data to a PC, and power supply and clock/reset modules to ensure system stability and synchronization. The physical layout of the platform is shown in Figure~\ref{实物图}.

In the RF pulse transmission section, a DAC chip is used to convert the digital pulse signals into analog form. These signals are subsequently amplified and filtered through amplifier and filter circuits to generate the RF pulse signals that drive the coil, producing the necessary magnetic field. To ensure high-quality RF pulse generation, the AD9767 chip, which operates at 125 MSPS with 14-bit resolution, is selected for digital-to-analog conversion.

In the magnetic resonance signal reception section, an ADC chip is used to acquire the magnetic resonance signals. Prior to acquisition, the signals are preprocessed by an analog front-end module, which includes a variable gain amplifier, band-pass filter, and low-pass filter to condition the received signals. Additionally, eight identical signal processing channels are designed to enable simultaneous parallel reception of eight-channel signals, enhancing the SNR and accelerating the data acquisition process. The AD9257 chip, with 65 MSPS and 14-bit resolution, is chosen to meet the required sampling rate and digital resolution. This chip supports synchronous processing of eight channels on a single chip, fulfilling performance requirements and improving system integration.

To accurately transmit the processed signals to a PC, the DDR3 module is used for intermediate data storage, and an Ethernet interface is provided for data transmission. Additionally, signal integrity and power integrity analyses are performed to ensure the quality of the transmitted signals.

\subsection{Embedded Digital System}

\begin{figure}[h]
  \centering
  \includegraphics[width=0.5\textwidth]{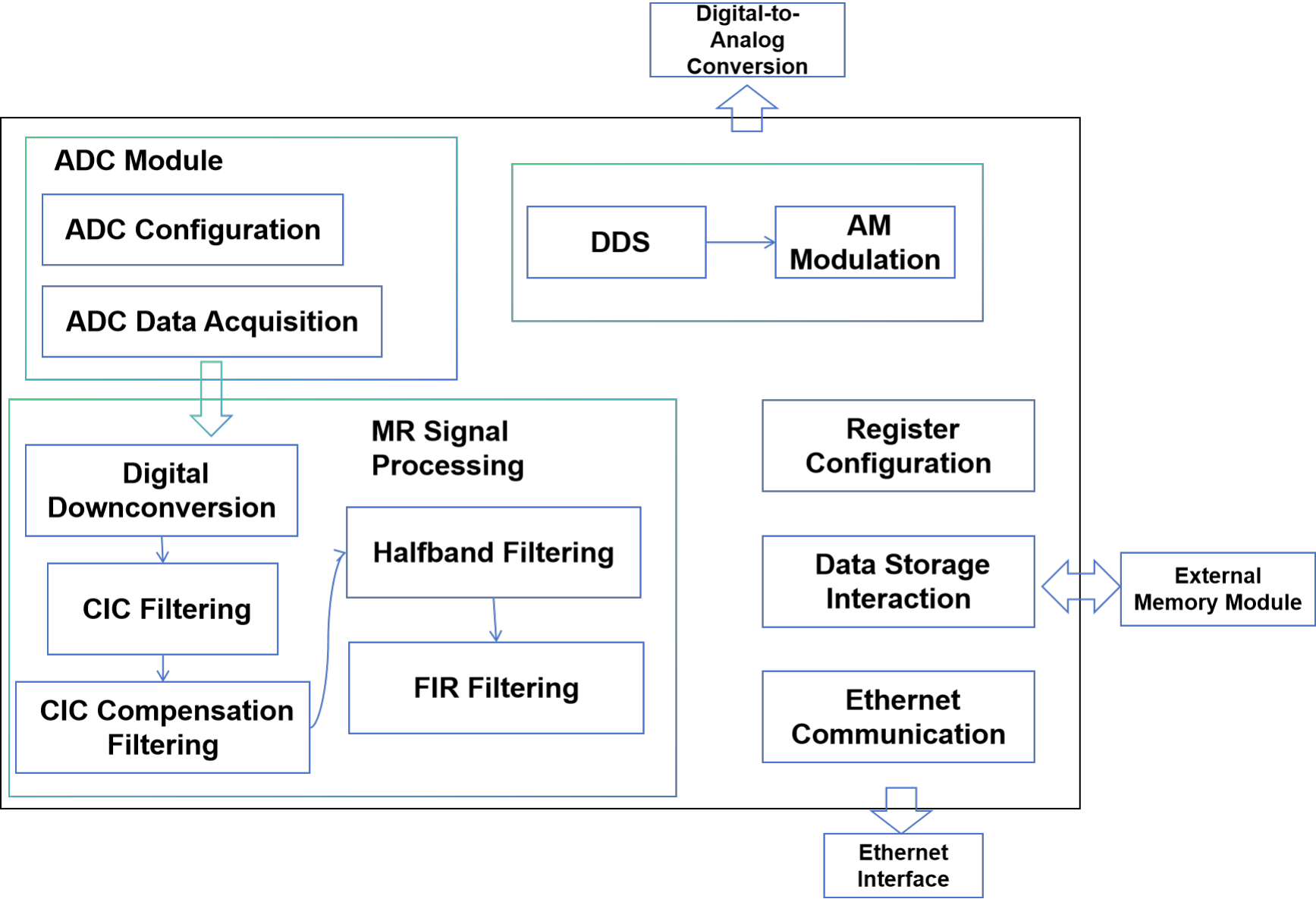}
  \caption{The Framework of Digital System.}
  \label{Digital System}
\end{figure}

The digital system is implemented based on FPGA and hardware logic. Its detailed architecture is illustrated in Figure~\ref{Digital System}, and includes the RF pulse transmission section, the magnetic resonance signal processing section, the ADC interaction module, the register configuration module, the storage interaction module, and the Ethernet interface module.

For RF pulse generation, based on the DDS architecture described earlier, the functionality implemented within the FPGA primarily includes frequency registers, phase accumulators, phase registers, waveform memory, and multipliers. Specifically, the frequency register is 32 bits, the phase and amplitude registers are 14 bits each, and the size of the phase lookup table is 16,384 × 14 bits.

In detail, a data storage file in COE format needs to be generated using MATLAB. This file can create various waveform files, such as Gaussian waveforms and sinc waveforms, according to the requirements of the magnetic resonance RF pulse. The content of this file corresponds to the waveform memory of the sine lookup table. Each time the system clock triggers, a frequency tuning word is added to the phase accumulator, which continuously accumulates and combines with the initial phase value to produce a 32-bit phase value stored in the phase register. The highest 14 bits of the accumulated value are then used as the address for the waveform memory to look up the amplitude of the sine wave. This process completes the phase-to-amplitude mapping, with each readout forming a digital baseband waveform. After DAC conversion, the desired analog signal is obtained.

According to the DDS principle, the FPGA generates both a sinc wave and a sine wave, with modulation capabilities for frequency, amplitude, and phase for each. Within the FPGA, a multiplier combines these two signals at a specific frequency to perform amplitude modulation and produce the RF pulse signal. The detailed block diagram of the internal FPGA structure is shown in Figure~\ref{Digital System}.


For processing the acquired magnetic resonance signals, the DDS module generates orthogonal local oscillators that are coherent in both frequency and phase with the carrier of the input MRI signal, based on the parameters and control signals configured by the register configuration module. These coherent oscillators are multiplied with the input MRI signal using a multiplier to achieve coherent demodulation, resulting in two baseband signals: in-phase (I) and quadrature (Q). The baseband signals are then processed through a series of filters designed to remove out-of-band noise and reduce the data rate, thus achieving digital downconversion. Configuration parameters include control settings input to the register configuration module and filter coefficients designed in MATLAB.
For the magnetic resonance system operating with a 20 kHz bandwidth input MRI signal, the filters were simulated using MATLAB. Based on the frequency response curves, adjustments were made to the number of filter stages and the allocation of decimation factors for each stage to optimize performance. 

The digital downconversion module in this system is composed of a digital quadrature mixer, CIC filter, CIC compensation filter, half-band filter, and FIR filter, as illustrated in Figure~\ref{Digital System}. The process begins with generating a coherent quadrature local oscillator signal that is frequency and phase-aligned with the carrier of the input MRI signal using a DDS. This signal is then mixed with the input MRI signal to achieve coherent demodulation, producing in-phase (I) and quadrature (Q) baseband signals.

These baseband signals are subsequently processed through a series of filters to eliminate out-of-band noise and reduce the data rate, culminating in the generation of the baseband digital signal and completing the digital downconversion process.

The configuration of the module is governed by parameters including user-defined control settings input via an interactive interface and filter coefficients designed in MATLAB. For the MRI signal with a 20 kHz bandwidth, filters were simulated and optimized in MATLAB. By iteratively adjusting the number of filter stages and the decimation factors based on frequency response curves, the final filter architecture was established, as depicted in Figure~\ref{Digital System}. This architecture comprises a two-stage CIC filter, a two-stage CIC compensation filter, a three-stage half-band filter, and a single FIR filter. 
\section{Experiment Setup}
\begin{figure}[h]
  \centering
  \includegraphics[width=0.5\textwidth]{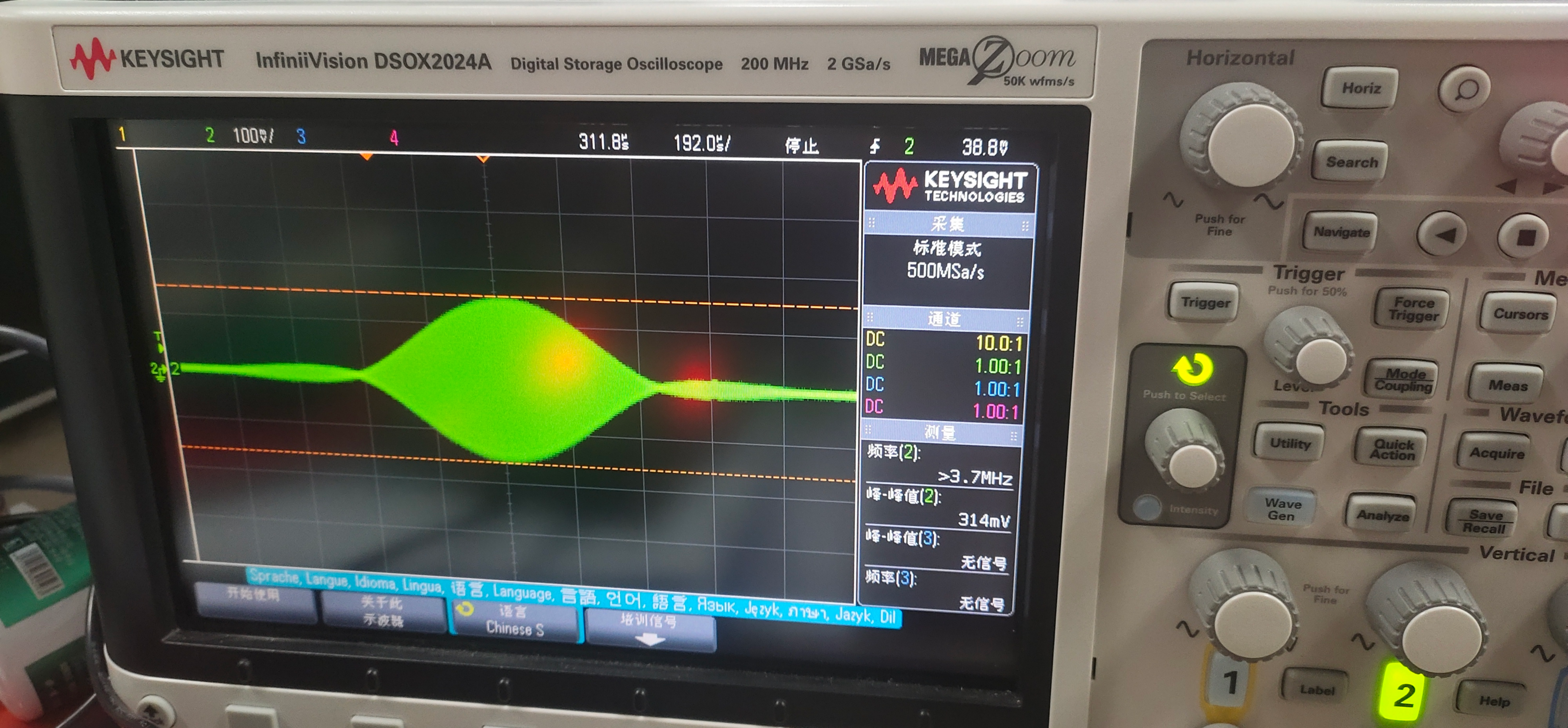}
  \caption{RF pulse waveform generated by the transmitter section.}
  \label{sinc}
\end{figure}

To validate the effectiveness of our platform, we performed a series of tests using simulation tools, oscilloscopes, and ultimately implemented the system on an MRI platform, achieving a range of high-performance results. Figure~\ref{sinc} shows the RF pulse generated by the transmitter section, which has been compared with the output from a commercial spectrometer. The RF pulse has a frequency of 13.88 MHz, an amplitude of 317 mV, and a duration of 2.58 ms, demonstrating successful production of the desired pulse.

\begin{figure}[h]
  \centering
  \includegraphics[width=0.5\textwidth]{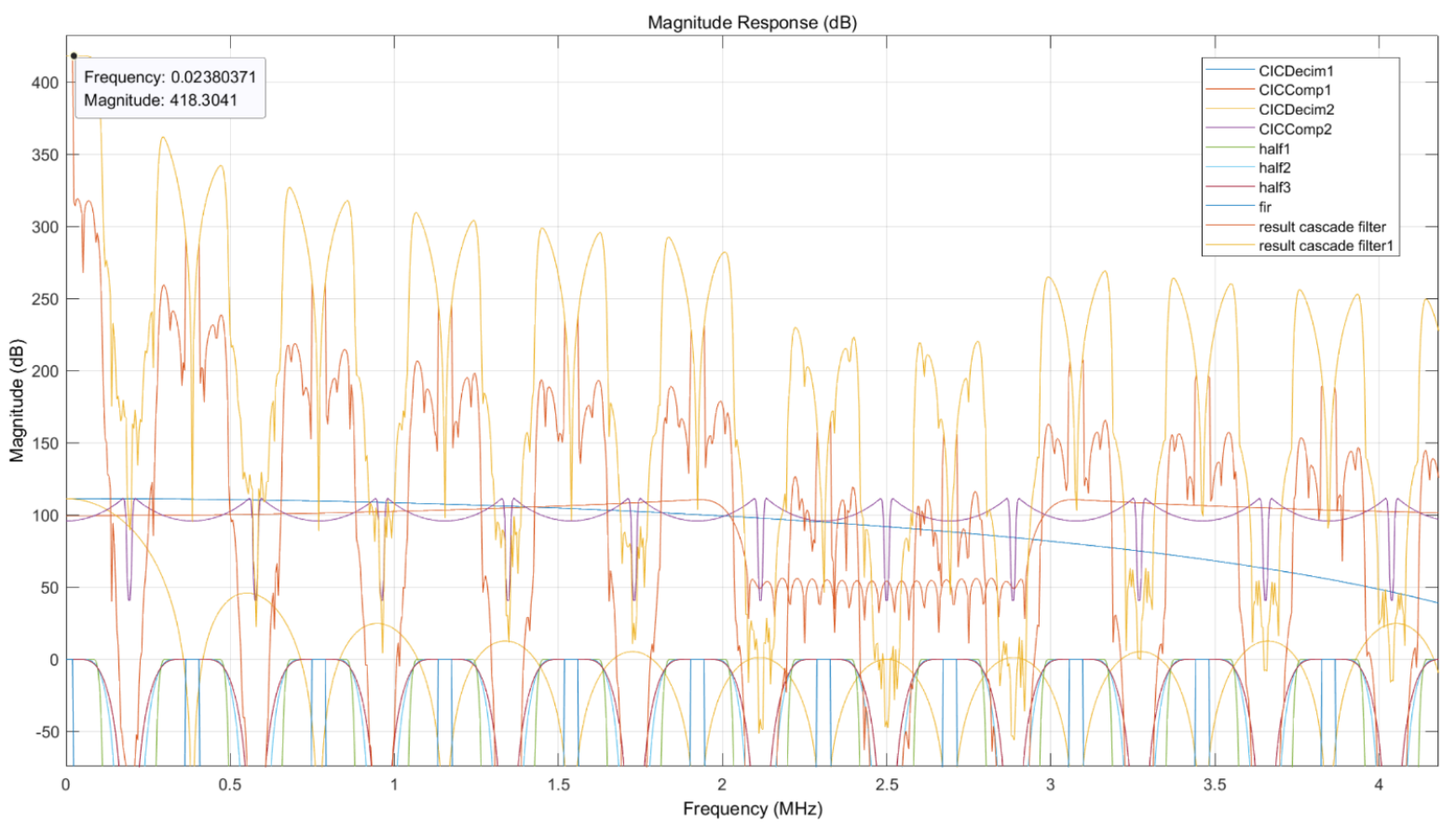}
  \caption{Amplitude-frequency response of multi-stage filter cascades.}
  \label{signalprocess}
\end{figure}

Figure~\ref{signalprocess} depicts the amplitude-frequency response of the filter after cascading different filters in the receiver section. This response was configured and obtained using MATLAB. Following satisfactory simulation results, the filter coefficients were exported as COE files and imported into Xilinx IP cores for FPGA implementation. From the Figure \ref{signalprocess}, it can be observed that the multi-stage cascaded filters ensure high-quality magnetic resonance signals within a 20 kHz bandwidth while performing decimation of the signal.

\begin{figure}[h]
  \centering
  \includegraphics[width=0.5\textwidth]{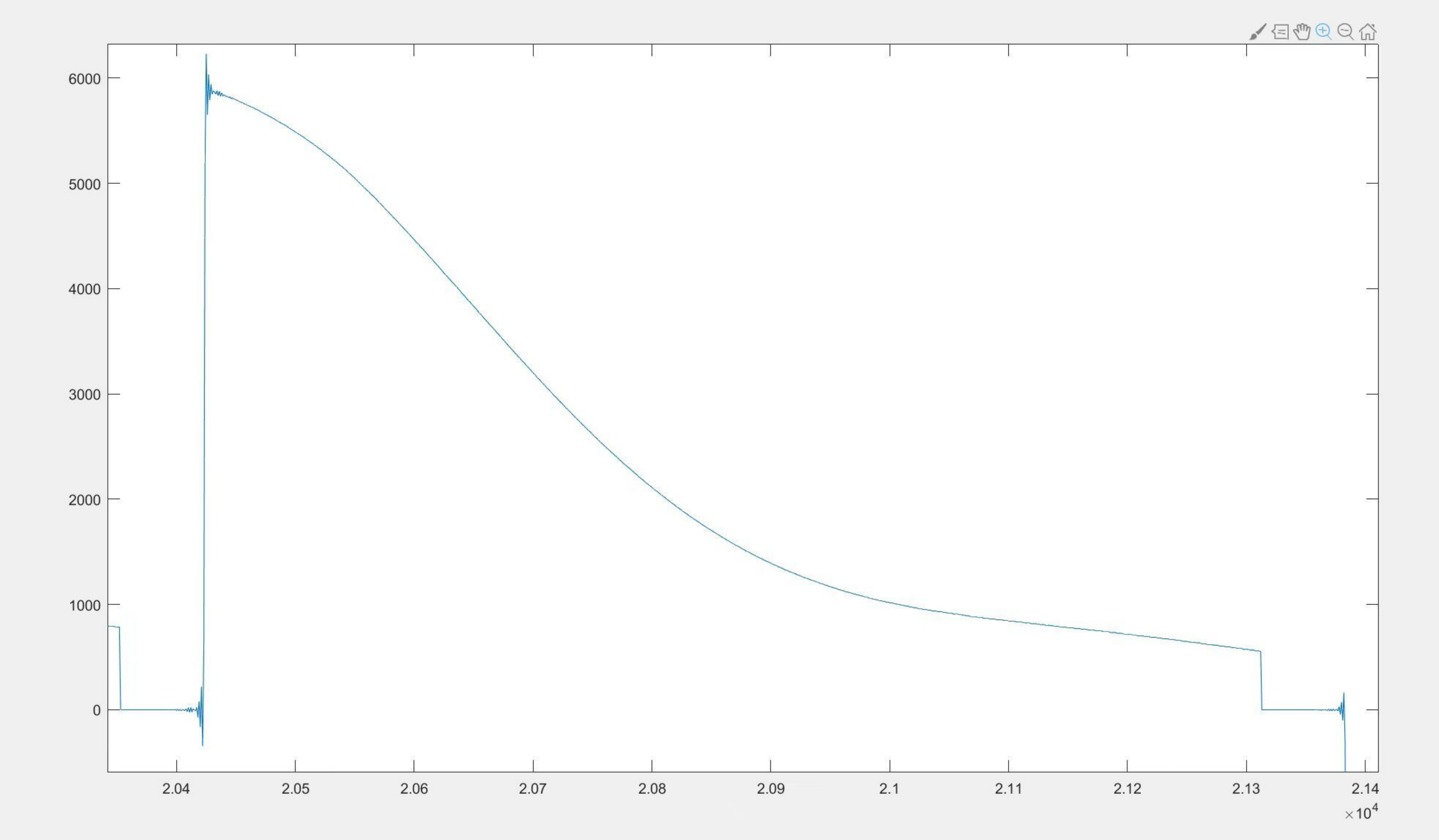}
  \caption{FID signal obtained through the MRI transmission and reception processing platform.}
  \label{fid}
\end{figure}

Figure~\ref{fid} shows the FID signal acquired through our MRI transmission and reception processing platform. The FID signal was obtained after transmitting an RF pulse to excite hydrogen nuclei in the body, followed by data acquisition and processing within the platform's receiver section. The quality of the FID signal is notably high, as evidenced by the results.

\begin{figure}[h]
  \centering
  \includegraphics[width=0.5\textwidth]{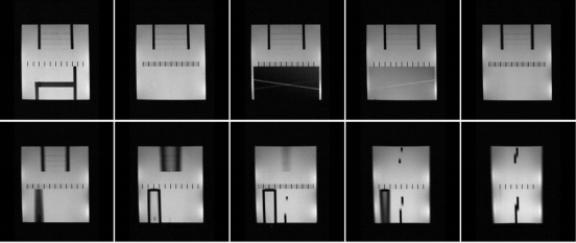}
  \caption{Phantom image obtained through our MRI platform.}
  \label{phantom}
\end{figure}

Figure~\ref{phantom} presents a phantom image obtained through our platform, demonstrating that the platform can replace the MRI spectrometer in both RF pulse transmission and magnetic resonance signal acquisition and processing, ultimately producing clear and detailed images.

\section{Conclusion}
This paper introduces an FPGA-based platform designed for RF pulse transmission and MRI signal reception and processing. Experimental results confirm that the platform effectively fulfills its intended functions and performs robustly within an MRI system. We intend to make this platform open-source on GitHub, aiming to encourage broader collaboration and development in the field of MRI spectrometers.
\bibliographystyle{IEEEtran}
\bibliography{ICICSP}
\end{document}